systematic quality assurance and management support.

## 2. Software Project Estimation

SPRINT I can be applied to several attributes of interest (e.g., effort or defect distribution). For clarity, in the following description only one attribute is considered. The technique can be sketched as follows: Measurement data from former projects are analyzed in order to build up characteristic curves (one for each project), which are comparable to each other. Afterwards, clusters of projects are identified. Based on the context of the project to be planned, the technique selects a suitable cluster and uses its cluster curve (mean of all characteristic curves within a cluster) for prediction. During the enactment of the project, the prediction is adapted with respect to actual project data. This leads to an empirical-based prediction and to flexibility for project and context changes.

The first step is to compute a characteristic curve for each of the previous projects in the database. These curves are used to make the projects comparable to each other. For instance, we compute the project's effort in percentage out of absolute person hour values. All characteristic curves need the same granularity; that is, we have to smooth values of the original series, or interpolate values.

The second step is to create clusters of similar characteristic curves. We use, for instance, the Euclidian distance to identify these curves and compute the mean of all curves within a certain cluster afterwards. This curve is called cluster curve. A problem is to determine the context of a single cluster (i.e., the scope of validity of the cluster curve), for the context knowledge of all included characteristic curves has to be considered. So, which context values are typical for a certain attribute of the cluster, like the "complexity" of the resulting system. Very difficult to handle are nominal scale values, because it is not possible to compute a mean. In this case, we compute the frequency of meeting a value in the characteristic curves of the cluster and in the database of all curves. In Figure 1, we have four possible values for "complexity". Three of them, namely "C", "D", and "E" are part of the context knowledge of the curves included in the cluster. Next, we have to compute a weight for each of the included values, determining the value's significance. Therefore we compute the relation of both frequencies and divide the result by a norm of all possible values to make the resulting

## Abstract

*Using quantitative data from past projects for software project estimation requires context knowledge that characterizes its origin and indicates its applicability for future use. This article sketches the SPRINT I technique for project planning and controlling. The underlying prediction mechanism is based on the identification of similar past projects and the building of so-called clusters with typical data curves. The article focuses on how to characterize these clusters with context knowledge and how to use context information from actual projects for prediction. The SPRINT approach is tool-supported and first evaluations have been conducted.*

## 1. Introduction

An essential task during software project planning is the identification of target values for key factors (such as effort, cost, schedule). This offers opportunities for reacting early and preventing plan deviations. The fundamental project planning activity for getting such target values is estimation. Precise estimates are essential for better project planning and control, and can lead to successful project delivery. Getting accurate estimation models for software development projects is a difficult task. The underlying processes are not unique and have to be adapted to project constraints and organizational constraints. Thus, estimation models need to be adaptable to specific project and organizational contexts (such as tool environment, application domain). Practical experience has shown that in many software-developing organizations, data from past projects is available, which could be used for predictions. This article proposes a technique called SPRINT I [3], which uses such data from past projects for estimations. The technique is based on cluster analysis introduced by Li and Zelkowitz in 1993 [1]. Cluster analysis is a means for finding groups in data that represent the same information model (e.g., similar effort distributions). A subsequent analysis of the corresponding context characteristics of the projects in one group supports the assignment of actual projects to similar past projects and helps to predict the future behavior more accurately. SPRINT I was developed in the context of a software project control center (SPCC) [2], a means for weights comparable to each other. Next, we have to choose a threshold that determines whether a value of a certain weight is typical for the cluster. In Figure 1, we chose all positive values; that is, in case of "complexity" value "D" of weight "0.27" and "E" of weight "0.95.

The third step is to assign a project to one of the computed clusters estimating a certain project attribute (like effort or costs). Because the project itself has yet no distribution for this attribute, we use the context knowledge to find a suitable cluster. Therefore we compute the "goodness of fitting" to a certain cluster by simply adding the computed weights for each context value that appears in the cluster as well as in the context of the project itself. For example, if a project has domain "B" and complexity "D" in its context, the goodness of fitting to the cluster of Figure 1 is "0.97 + 0.27 = 1.24". We assign the project to the cluster of the highest goodness of fitting value and use the cluster curve to estimate the attribute's distribution.

The fourth step takes place during project execution. Actual values of an attribute are compared to estimated values. If the distance between the two is above (or below) a most tolerable threshold, project management has to be informed in order to initiate replanning steps. In our case, this means to dynamically find a new cluster in order to predict the project's behavior in a better way. The deviation cause can be manifold. For instance, the project's initial context could possibly be wrong or was changed because of internal problems. The first assignment of the project to a certain cluster was just based on context knowledge. Now, we additionally consider the attribute's actual distribution; that is, we use both information to find a suitable cluster: On the one hand, we use the Euclidian distance to compute the similarity between the actual project curve and a cluster curves. On the other hand, we use the context knowledge to identify a suitable cluster. Several strategies exist to choose a new estimation cluster, for instance, minimal Euclidian distance, maximal goodness of fitting, or a combination of both, like Euclidian distance less a certain threshold and maximal goodness of fitting at the same time.

## 3. Evaluation

The core technique has been implemented in an SPCC prototype tool at the University of Kaiserslautern and has been applied to a set of sample projects. Currently, the user has to choose manually what kind of strategy he or she wants to use to assign a project to a cluster. We basically provide three kinds of strategies: Euclidian distance, goodness of fitting, and hybrid. In the latter case, we try to find an optimal combination between the first two strategies. First results show small average deviations between predicted and actual values.

## 4. Related Work and Conclusion

Clustering techniques can be roughly classified as follows: a) *Hierarchical techniques*: The clusters themselves are classified into groups, the process being repeated at different stages to form a tree. b) *Optimization-partitioning techniques*: Clusters are formed by optimization of a so-called "clustering criterion". The clusters are pair-wise exclusive, as a result forming a partition of the set of entities. c) *Density or mode-seeking techniques*: Here, clusters are formed by looking for regions that contain a relatively dense concentration of entities. This technique tends to include entities into existing clusters rather than to initiate new clusters. d) *Clumping techniques*: Here, the clusters or clumps can overlap. These techniques can be used in some cases (e.g., language studies), where an overlap between classes is allowed (e.g., words tend to have several meanings). In SPRINT I the prediction and control of project progression is directly based on the experience in past projects, the accuracy of planned curves is increased by data- and context-driven selection of cluster curves, the approach is directly applicable without a number of reference applications, and the adaptation of planned curves takes place dynamically.

*Acknowledgements*. This work was partly funded by the German Federal Ministry of Education and Research (BMBF) as part of SEV and the DFG as part of SFB 501.

## 5. References

[1] N.R. Li, M.V. Zelkowitz. *An Information Model for Use in Software Management Estimation and Prediction*. CIKM 1993, Washington, DC, USA, 1993, pp. 481-489.

[2] J. Münch, J. Heidrich. *Software Project Control Centers: Concepts and Approaches*. Journal of Systems and Software 70 (1).

[3] J. Münch, J. Heidrich A. Daskovska. *A Practical Way to Use Clustering and Context Knowledge for Software Project Planning*. SEKE 2003, San Francisco Bay, USA, 2003.

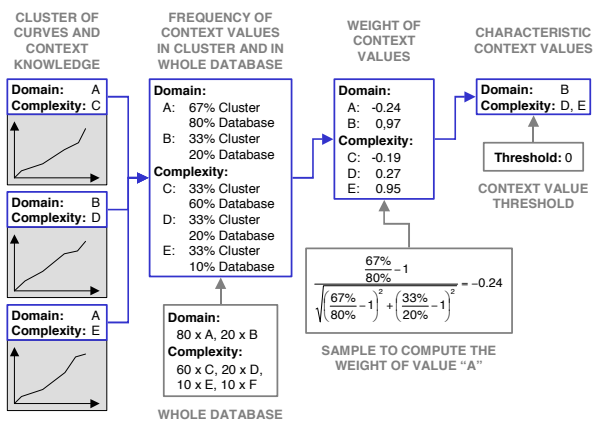

**Figure 1.** Determining context knowledge.